\documentclass{article}
\usepackage[utf8]{inputenc}
\usepackage{amsfonts,amsmath,amssymb}
\usepackage{bm}
\usepackage{booktabs}
\usepackage{mathrsfs}
\usepackage{bm}
\usepackage{graphicx}
\usepackage{caption}
\usepackage{subcaption}
\usepackage{url}
\usepackage{hyperref}
\hypersetup{colorlinks=false,pdfborder={0 0 0}}
\usepackage{cleveref}

\ifpdf
\hypersetup{
  pdftitle={PF-EnKF Hybrid},
  pdfauthor={G. Robinson and I. Grooms}
}
\fi

\title{A hybrid particle-ensemble Kalman filter for problems with medium nonlinearity\thanks{This work was supported by the US National Science Foundation under grant no. DMS-1821074. This work utilized the Summit supercomputer, which is supported by the National Science Foundation (awards ACI-1532235 and ACI-1532236), the University of Colorado Boulder, and Colorado State University. The Summit supercomputer is a joint effort of the University of Colorado Boulder and Colorado State University.}}
\author{Gregor Robinson \and Ian Grooms}
\date{Department of Applied Mathematics, University of Colorado, Boulder}

\begin{document}

\maketitle 
\begin{abstract}
A hybrid particle ensemble Kalman filter is developed for problems with medium non-Gaussianity, i.e. problems where the prior is very non-Gaussian but the posterior is approximately Gaussian. Such situations arise, e.g., when nonlinear dynamics produce a non-Gaussian forecast but a tight Gaussian likelihood leads to a nearly-Gaussian posterior. The hybrid filter starts by factoring the likelihood. First the particle filter assimilates the observations with one factor of the likelihood to produce an intermediate prior that is close to Gaussian, and then the ensemble Kalman filter completes the assimilation with the remaining factor. How the likelihood gets split between the two stages is determined in such a way to ensure that the particle filter avoids collapse, and particle degeneracy is broken by a mean-preserving random orthogonal transformation. The hybrid is tested in a simple  two-dimensional (2D) problem and a multiscale system of ODEs motivated by the Lorenz-`96 model. In the 2D problem it outperforms both a pure particle filter and a pure ensemble Kalman filter, and in the multiscale Lorenz-`96 model it is shown to outperform a pure ensemble Kalman filter, provided that the ensemble size is large enough.
\end{abstract}

\section{Introduction}
Data assimilation of high-dimensional dynamical systems routinely falls to various kinds of ensemble Kalman filters (EnKF) \cite{Evensen09}.
Ensemble Kalman filters make two fundamental approximations: the first is that the likelihood and prior are both Gaussian, and the second is that the mean and covariance of the Gaussian prior are approximated from an ensemble.
The EnKF is known to converge to the correct posterior in the limit of large ensemble size when the distributions are Gaussian \cite{MCB11}, but it clearly will not converge to the correct posterior in the presence of non-Gaussianity.

In contrast, Sequential Importance Sampling with Resampling (SIR a.k.a. Particle Filtering) is known to weakly converge to the correct posterior in the large-ensemble limit --- with remarkably mild constraints on the dynamics, prior, and observing system \cite{GSS93,CD02,LSZ15}.
This flexibility makes SIR superficially attractive for applications like weather forecasting where nonlinear fluid dynamics lead to non-Gaussian distributions.
Unfortunately, however, SIR suffers a severe curse of dimensionality that has prevented its practical application to high dimensional data assimilation problems \cite{BBL08,SBBA08,SBM15}.
A variety of methods have been proposed to improve the performance of particle filters in high-dimensional problems, including implicit particle filters \cite{CT09,CT12,CMT10}, the equivalent-weights particle filter \cite{vanLeeuwen09,AvL13,AvL15,ZvLA15,SEvLA19}, likelihood approximations \cite{RGK18}, local particle filters \cite{RvH15,PM16,Poterjoy16} and particle filters based on kernel mappings \cite{PvL19} and synchronization methods \cite{PvLG19}.
Particle filters have also been hybridized with EnKFs \cite{FK13,CRR16,PMCG10,SSAS15} and with variational methods \cite{MHP18}.
Methods have also been proposed to mitigate the assumption of Gaussianity within the Kalman filter, including nonlinear transformations on the univariate marginal distributions (termed `Gaussian anamorphosis' in the literature \cite{BEW03,BPW10,ZGFL11,BrankartEtAl12,SB12}) and methods based on rank statistics \cite{Anderson10,MCSB14,Anderson19,Anderson20}.

Although nonlinear dynamics, nonlinear observation operators, and non-Gaussian error distributions lead to non-Gaussian priors and likelihoods in many applications, the degree of non-Gaussianity is not always so great that it severely degrades EnKF performance.
This has led several authors to classify problems according to the degree of nonlinearity, i.e. the degree of non-Gaussianity \cite{Bocquet11,MCSB14,MH19}.
Following \cite{MH19} we distinguish three categories:
\begin{itemize}
    \item Mild nonlinearity: The prior and posterior are both approximately Gaussian.
    \item Medium nonlinearity: The prior is very non-Gaussian but the posterior is approximately Gaussian.
    \item Strong nonlinearity: The prior and posterior are both very non-Gaussian.
\end{itemize}
Particle filters and non-Gaussian extensions of the EnKF are not needed in situations with mild nonlinearity, while problems with strong nonlinearity can greatly benefit from such methods.
Problems with medium nonlinearity can arise when nonlinear dynamics produce a non-Gaussian prior, but a highly accurate Gaussian likelihood generates a nearly Gaussian posterior.
The concept of medium nonlinearity is related to the Laplace approximation \cite{TK86}.
Morzfeld and Hodyss \cite{MH19} argue that variational methods are more appropriate for medium nonlinearity than EnKF methods because the former make a Gaussian approximation of the posterior, while the latter make a Gaussian approximation of the prior.

The goal of the present work is to develop a hybrid of the SIR particle filter with the EnKF that is appropriate for problems with medium nonlinearity.
The hybrid is based on the likelihood splitting of Frei \& K\"unsch \cite{FK13}. 
At each assimilation cycle, part of the observational information is incorporated by means of an SIR step, and then the remaining observational information is incorporated with a serial square root version of the EnKF.
Particle degeneracy that results from the resampling step of the SIR is broken by a mean preserving random orthogonal transformation of the ensemble, as seen in certain EnKFs \cite{Evensen04,LEB05,SO08} and moment-matching particle filters \cite{LB11,TA15}.
The goal of the hybrid is to present the EnKF with an intermediate prior that is closer to Gaussian than the true prior.
The curse of dimensionality in the particle filter is mitigated by assimilating only part of the observational information, i.e. only moving partway from the prior to the posterior, thereby enabling accurate results with practical ensemble sizes.
The hybrid presented here is broadly similar to other hybrids (e.g. \cite{FK13,CRR16}), and differs mainly in the explicit focus on problems with medium nonlinearity and in details of the implementation.
Differences are discussed further in section \ref{sec:Hybrids}.

The hybrid particle ensemble Kalman filter is presented in section \ref{sec:Hybrid}.
The new hybrid is compared to the hybrids from \cite{FK13,CRR16} and to a particle filter and an ensemble Kalman filter in the context of a simple two-dimensional problem in section \ref{sec:Henon}.
A multiscale Lorenz-'96 model from \cite{GL15} is described in section \ref{sec:L96}, followed by a description of the data assimilation system configuration in section \ref{sec:DASetup}.
The EnKF component of the hybrid uses multiplicative inflation and localization, and the method used to optimize the values of these parameters is described in section \ref{sec:GP}.
Results of the experiments are described in section \ref{sec:Results}, followed by a conclusion in section \ref{sec:Conclusion}.

\section{The hybrid algorithm}
\label{sec:Hybrid}

\subsection{SIR}
Standard sequential importance resampling (SIR) particle filters work as follows \cite{GSS93,DdFG01}.
Each ensemble member $\bm{x}_0^{(i)}$ (or `particle') starts with equal weight $w_0^{(i)} = 1/N$, where $N$ is the ensemble size and $i=1,\ldots,N$.
Subscripts refer to time, while superscripts in parenthesis refer to ensemble members.
Each ensemble member is forecast until the next assimilation cycle.
At assimilation cycle $j$ the weights are updated using the likelihood $L(\bm{x})$
\begin{equation}
w_j^{(i)} = \tilde{w}_{j-1}^{(i)}\frac{L\left(\bm{x}_j^{(i)}\right)}{Z_j}
\end{equation}
where $Z_j$ is a normalization constant to ensure that the weights sum to one, and $\tilde{w}_j$ denotes the effect of resampling: without resampling at step $j$ we have $\tilde{w}_j^{(i)} = w_j^{(i)}$ whereas with resampling we have $\tilde{w}_j^{(i)} = 1/N$.
A resampling is then applied whereby particles with high weights are replicated and particles with low weights are eliminated.
There are a variety of resampling algorithms; here we use the so-called `systematic' resampling scheme of \cite{Kitagawa96}.

It is well known that the weights of a particle filter can collapse, especially in high dimensions, i.e. a small number of particles receive a weight near one while all others receive a weight near zero \cite{SBBA08,SBM15}.
After resampling, only the high-weight particles are left.
If an optimal-transport based alternative to resampling is used \cite{Reich13,CRR16,AdWR17}, then all particles are transported to a very small vicinity of the high-weight particles.
In both cases the posterior distribution is poorly estimated.
The number of particles with a substantial portion of the weight can be approximated by the effective sample size
\begin{equation}
\text{ESS} = \frac{1}{\sum_{i=1}^N\left(w_j^{(i)}\right)^2}.
\end{equation}
The ESS takes values between 1 and $N$, and small ESS indicates that the weights have collapsed.

\subsection{Ensemble Square Root Filter (ESRF)}\label{sec:ESRF}
There are many ensemble Kalman filters, any of which could be hybridized with the smoothed particle filter.
We focus here on an ensemble square root filter (ESRF) developed in \cite{WH02} for sequential assimilation of observations possessing uncorrelated errors.
At a single assimilation cycle the ensemble is denoted $\{\bm{x}^{(i)}\}_{i=1}^N$.
The ensemble mean is denoted $\bm{\bar{x}}$, and the scaled ensemble perturbation matrix is denoted
\begin{equation}
\mathbf{A} = \frac{1}{\sqrt{N-1}}\left[\bm{x}^{(1)}-\bar{\bm{x}},\ldots,\bm{x}^{(N)}-\bar{\bm{x}}\right].
\end{equation}
The ensemble covariance matrix is thus $\mathbf{AA}^T$.
Covariance inflation is applied by replacing $\mathbf{A}$ with $\sqrt{1+r}\mathbf{A}$, where $r>0$ is a tunable inflation factor.

Observations are linear, and a single scalar observation $y$ takes the form
\begin{equation}
y = \bm{h^Tx} + \epsilon.
\end{equation}
Here the observation error $\epsilon$ is a sample from a zero-mean normal distribution with variance $\gamma^2$ and the row vector $\mathbf{H} = \bm{h}^T$ extracts the observations from the state vector $\bm{x}$.
It is convenient to define the row vector $\bm{v}^T = \bm{h}^T\mathbf{A}$.
With this notation, the ESRF from \cite{WH02} corresponds to the following update of the ensemble mean
\begin{equation}\label{eqn:ESRF_Mean}
\bar{\bm{x}}^a = \bar{\bm{x}} + \frac{(y-\bm{h}^T\bar{\bm{x}})}{\sigma^2+\gamma^2}\mathbf{A}\bm{v}
\end{equation}
and the following update of the scaled ensemble perturbation matrix
\begin{align}\label{eqn:ESRF_A}
\mathbf{A}^a &= \mathbf{A} - b\mathbf{A}\bm{vv}^T, \\
b &= \frac{1}{\sigma^2+\gamma^2+\gamma\sqrt{\sigma^2+\gamma^2}}
\end{align}
where $\sigma^2 = \bm{v}^T\bm{v}$ and $\mathbf{A}^a$ is the scaled analysis ensemble perturbation matrix.

Localization is applied by multiplying the increments elementwise by a localization vector $\bm{\rho}$.
The elements of $\bm{\rho}$ are $e^{-(d/L)^2/2}$, where $d$ is the distance from $\bm{x}_i$ to $y$ and $L$ is a tunable localization radius.
This amounts to updating Eq \ref{eqn:ESRF_Mean} and Eq \ref{eqn:ESRF_A} to
\begin{equation}\label{eqn:ESRF_MeanL}
\bar{\bm{x}}^a = \bar{\bm{x}} + \frac{(y-\bm{h}^T\bar{\bm{x}})}{\sigma^2+\gamma^2}\bm{\rho}\circ\left(\mathbf{A}\bm{v}\right)
\end{equation}
and
\begin{equation}\label{eqn:ESRF_AL}
\mathbf{A}^a = \mathbf{A} - b\left(\bm{\rho}\circ\left(\mathbf{A}\bm{v}\right)\right)\bm{v}^T
\end{equation}
where $\circ$ denotes an elementwise product.

Evensen was the first to suggest resampling the posterior within the context of an ensemble square root filter by multiplying $\mathbf{A}^a$ from the right by a random orthogonal matrix \cite{Evensen04}.
Since the posterior ensemble covariance matrix is $\mathbf{A}^a\mathbf{A}^{aT}$, this kind of resampling does not change the ensemble covariance matrix.
Sakov \& Oke \cite{SO08} pointed out that the random orthogonal matrix should have $\bm{1}$ (the vector whose elements are all $1$) as an eigenvector in order for the resampling to preserve the ensemble mean.
We construct a new scaled ensemble perturbation matrix $\mathbf{A}^a$ by multiplying $\mathbf{A}^a$ from the right by a random orthogonal matrix $\mathbf{Q}$ that has $\bm{1}$ as an eigenvector.
The matrix $\mathbf{Q}$ is constructed as follows \cite{SO08}
\begin{equation}
\mathbf{Q} = \mathbf{U}\left[\begin{array}{c|c}1&\bm{0}\\
\hline
\bm{0}&\mathbf{P}\end{array}\right]\mathbf{U}^T.
\end{equation}
The matrix $\mathbf{U}$ is an orthogonal matrix whose first column is proportional to $\bm{1}$, while the matrix $\mathbf{P}$ is a random orthogonal matrix of size $N-1\times N-1$.
The matrix $\mathbf{U}$ is time independent.
With a large ensemble size it can become costly to sample a new $\mathbf{P}$ at each assimilation cycle. In principle the matrix $\mathbf{Q}$ could be constructed once and used repeatedly,
but in our numerical experiments $\mathbf{P}$ is resampled at each assimilation cycle.

Using this method, a single assimilation cycle proceeds as follows
\begin{itemize}
    \item Form the ensemble mean $\bar{\bm{x}}$ and scaled ensemble perturbation matrix $\mathbf{A}$.
    \item Inflate the scaled ensemble perturbation matrix: $\mathbf{A}\leftarrow (1+r)\mathbf{A}$
    \item For each observation, find $\bar{\bm{x}}^a$ and $\mathbf{A}^a$ using Eq \ref{eqn:ESRF_MeanL} and Eq \ref{eqn:ESRF_AL}.
    \item Resample the posterior ensemble by replacing $\mathbf{A}^a$ with $\mathbf{A}^a\mathbf{Q}$.
    \item Reconstitute the ensemble according to $\bm{x}^{(i)} = \bar{\bm{x}}^a+\sqrt{N-1}\mathbf{A}_i$ where $\mathbf{A}_i$ is the $i^\text{th}$ column of $\mathbf{A}$.
\end{itemize}

\subsection{SIR-ESRF hybrid}\label{sec:Hybrids}
The SIR/ensemble square root filter (SIR-ESRF) hybrid developed here is based on the bridging method of Frei and K\"unsch \cite{FK13}.
The likelihood $L(\bm{x})$ is split into a product $(L(\bm{x}))^\alpha \cdot (L(\bm{x}))^{1-\alpha}$ where $\alpha\in[0,1]$ is the ``splitting factor".
The hybrid proceeds by having the SIR particle filter assimilate using the likelihood $(L(\bm{x}))^\alpha$, followed by an ESRF assimilation using the likelihood $(L(\bm{x}))^{1-\alpha}$.
In principle, the methods can be applied in either order \cite{CRR16}, but the method developed here is intended for situations where the prior is non-Gaussian but the posterior is nearly Gaussian (`medium' nonlinearity according to \cite{MH19}).
In such cases the intermediate posterior produced after the first assimilation with the particle filter should be closer to Gaussian than the prior.
The ESRF subsequently performs an assimilation on a problem that more closely conforms to its underlying Gaussian approximation.

Following Frei \& K\"unsch \cite{FK13} we choose the splitting factor $\alpha$ to ensure that the effective sample size is within some tolerance of a tunable theshold.
This is achieved with a rootfinding method.
A large ESS threshold implies a small $\alpha$, though the precise value of $\alpha$ depends on the ensemble size.
If $\alpha=0$, then the hybrid reverts to a pure ESRF because all the particle filter weights become equal.

The resampling step of the SIR particle filter leads to a degeneracy where there are multiple copies of some ensemble members.
In our numerical experiments we use a deterministic system of ordinary differential equations, so the dynamics do not break the degeneracy.
We opt to follow the ESRF assimilation with a mean-preserving random orthogonal transformation that resamples the ensemble within the Gaussian posterior, as described in the foregoing section.

There are two other extant hybrid particle/ensemble Kalman filters: those of \cite{FK13} and \cite{CRR16}.
Our hybrid is essentially the same as the hybrid of \cite{CRR16} with the following differences: We use standard resampling methods for the particle filter part of the hybrid instead of the Ensemble Transform Particle Filter (ETPF) method of \cite{Reich13}, and we break degeneracy using a random orthogonal transformation rather than the `particle rejuvenation' procedure of \cite{CRR16}.
Our use of a random orthogonal transformation is motivated by the focus on medium nonlinearity problems.
Naive implementations of the ETPF are computationally expensive, and in the experiments with the H\'enon map described in section \ref{sec:Henon} there seems to be little benefit in using the ETPF instead of standard resampling.

The hybrid of \cite{FK13} is significantly different from the one proposed here and from the hybrid of \cite{CRR16} because the first step of Frei \& K\"unsch's hybrid is really a Gaussian mixture model update and not a particle filter update (cf. \cite{AA99}), though it does limit to a pure SIR particle-filter update in the limit $\alpha\to1$.
In particular the particle weights in the hybrid of \cite{FK13} are different from those used here and in \cite{CRR16}, and are more expensive to evaluate.
Particle degeneracy is avoided in the hybrid of \cite{FK13} by using a stochastic update for each step: a perturbed-observation Gaussian mixture update (cf. \cite{AA99}) for the first step and a perturbed-observation EnKF for the second step (cf. \cite{BvLE98,HM98}).
It is worth noting that the hybrids of \cite{FK13} and \cite{CRR16} are generally intended to overcome non-Gaussianity in the filtering problem.
The hybrid developed here is quite similar to that of \cite{CRR16} but has a tighter focus: we expect the hybrid to achieve near-optimal performance on problems with medium nonlinearity, but not on problems with strong nonlinearity.

\subsection{Blurring observations}
The development of particle filters that avoid or reduce the incidence of collapse is an active area of research.
The authors recently proposed an alternative that uses the same forecast as the standard particle filter, but imposes a generalized random field model of observation errors \cite{RGK18}.
When the observation errors are Gaussian, the likelihood takes the form
\begin{equation}
L(\bm{x})\propto \text{exp}\left\{-\frac{1}{2}(\bm{y}-\bm{H}(\bm{x}))^T\mathbf{R}^{-1}(\bm{y}-\bm{H}(\bm{x}))\right\}.
\end{equation}
Here $\bm{y}$ is the observation vector, $\bm{H}$ is the observation (or `forward') operator, and $\mathbf{R}$ is the observation error covariance matrix.
In the particle filter of \cite{RGK18}, the observation error covariance matrix $\mathbf{R}$ is replaced by a covariance matrix that has increasing variance at small spatial scales.
In practice this is implemented by blurring (i.e. smoothing) the innovations $\bm{y}-\bm{H}(\bm{x})$.
The authors recently developed a fast algorithm for blurring scattered data in arbitrary dimensions for this purpose \cite{RG20}.

In the numerical experiments presented here, the spatial domain is periodic and Fourier methods are used to apply the blurring.
The true observation error covariance matrix is $\mathbf{R} = \gamma^2\mathbf{I}$.
In the particle filter with blurred observations this is replaced by $\gamma^2\left(\mathbf{S}^T\mathbf{S}\right)^{-1}$, where the matrix $\mathbf{S}$ corresponds to an operator that attenuates the Fourier coefficients using the following spectrum
\begin{equation}
\frac{1}{\left(1+ (\ell k)^2\right)^\beta}
\end{equation}
where $\beta$ and $\ell$ are tunable parameters and $k$ is the Fourier wavenumber.
More general blurring spectra are trivial to implement in our experiments, but the above blurring corresponds to the spectrum of the fast algorithm for scattered data developed in \cite{RG20}.

Replacing the true likelihood by a likelihood associated with spatial blurring means that the particle filter is approximating a distribution other than the true Bayesian posterior.
The effect of this blurring is to make the likelihood uninformative at small scales, so that the posterior reverts to the prior at small scales.
At large scales the blurring likelihood is close to the true likelihood, so the approximate posterior is close to the true posterior.
Blurring reduces the effective dimension of the problem by confining the dimensionality to that of the large scales.
This has the effect of reducing the minimum ensemble size needed to avoid collapse. 
It can also improve uncertainty quantification of large scales for a fixed ensemble size.

\section{Numerical experiment: H\'enon map}\label{sec:Henon}
This section serves to illustrate a specific problem with medium nonlinearity, and to compare the three hybrid particle/ensemble Kalman filters with a particle filter and an ensemble Kalman filter.
Rather than performing a cycled data assimilation experiment where the output of one cycle serves as the initial condition for the next,  we repeat the same experiment multiple times.
This serves to focus attention on a single Bayesian assimilation update, avoiding the complication associated with cycled data assimilation where the degree of non-Gaussianity can vary from one cycle to the next.

\begin{figure}
    \centering
   \includegraphics[width=\textwidth]{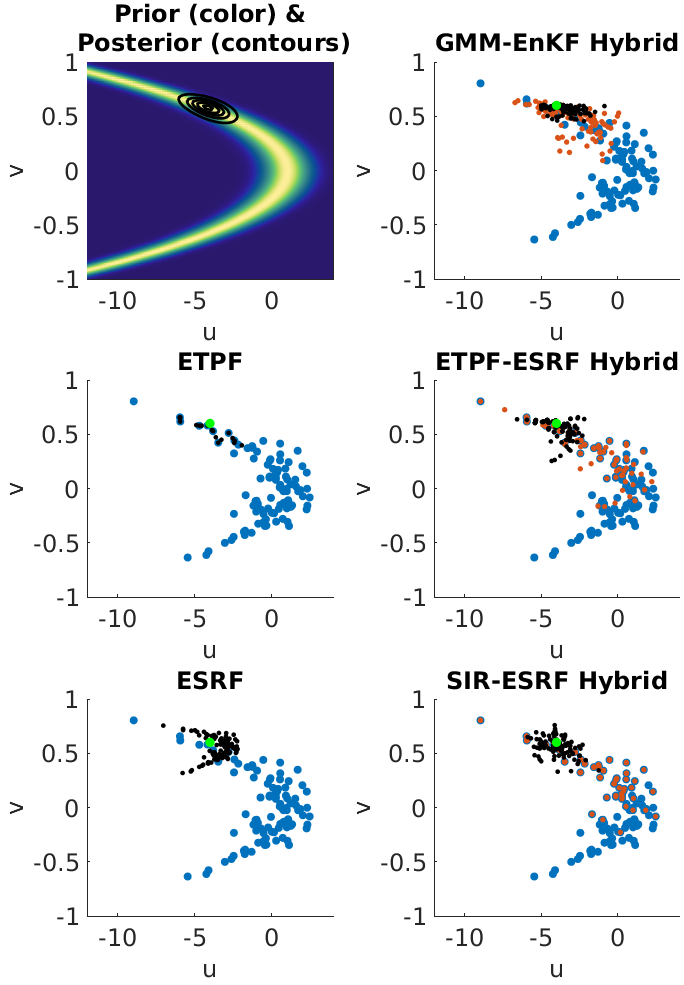}
    \caption{Upper left: Prior distribution (color) and posterior distribution (contours). The remaining panels illustrate the five methods. In each panel the blue dots show the prior ensemble, the black dots show the posterior ensemble, and the green dot shows the observation. In the right column, the orange dots represent the intermediate ensemble produced by the first step of each hybrid.}
    \label{fig:Henon}
\end{figure}

The prior imposed is the joint distribution of $U$ and $V$ obtained by applying one iteration of the H\'enon map to a standard normal initial condition on $U_0$ and $V_0$, i.e.
\begin{align*}
    U &= 1 - 1.4 U_0^2 + V_0,\\
    V &= 0.3U_0.
\end{align*}
The prior probability density is shown in color in the upper left panel of Fig \ref{fig:Henon}.
The true values of $U$ and $V$ are set to $-4$ and $0.6$, respectively, and the observation is drawn from the normal distribution with mean equal to the true value of $U$ and $V$ and diagonal covariance with entries $1$ and $0.01$.
The resulting posterior probability distribution is approximately Gaussian, as shown by the contours in the upper left panel of Fig \ref{fig:Henon} (where the observation is without error, for convenience).
Since the prior is clearly non-Gaussian, a pure EnKF solution is expected to give a biased result regardless of ensemble size.
In contrast, the hybrid should achieve nearly optimal performance provided that the ensemble size is large enough to avoid sampling errors.

To illustrate these ideas we compare five methods: 
\begin{itemize}
    \item[(i)] ETPF: A pure particle filter from \cite{Reich13}
    \item[(ii)] ESRF: A pure ESRF described in section \ref{sec:ESRF}
    \item[(iii)] GMM-EnKF: The Gaussian mixture model -- EnKF hybrid of \cite{FK13}
    \item[(iv)] ETPF-ESRF: The hybrid of \cite{CRR16} combining the ETPF and the serial square root ESRF described in section \ref{sec:ESRF}
    \item[(v)] SIR-ESRF: The hybrid described in section \ref{sec:Hybrids} that combines a standard SIR particle filter and an ESRF with a mean-preserving random orthogonal resampling
\end{itemize}
These five methods are illustrated in Fig \ref{fig:Henon}, using an ensemble size of 100; in every panel the blue dots represent the prior sample and the green dot shows the true value of $U$ and $V$.
The black dots represent the posterior ensemble in each panel, and in the panels illustrating the hybrid methods the orange dots represent the sample from the intermediate posterior distribution.

The center left panel illustrates the particle filter (ETPF).
The ETPF posterior sample is tightly clustered around a small number of the prior samples, which reflects the fact that the ESS is very low (ESS $=5$ in this example), despite having an ensemble size of 100 for a problem with dimension 2.
This illustrates the severe ensemble size requirements of particle filters.
The lower left panel shows the ESRF.
The ESRF produces a posterior close to the true value in this case, but the posterior ensemble it produces shows clear discrepancy from the true posterior.

The hybrids all choose the split $\alpha$ to produce an ESS of 30. 
ETPF-ESRF and SIR-ESRF are shown in the center right and lower right panels, respectively; they use the same split $\alpha$ and the same particle weights.
The two methods produce very similar results; one notable difference is that ETPF-ESRF produces an intermediate distribution with less particle degeneracy than SIR-ESRF.
This difference in the intermediate distribution does not have a significant impact on the final posterior distribution.
GMM-EnKF (upper right panel) uses a different formula for the particle weights --- because the first step is a Gaussian mixture model rather than a sum of delta distributions --- and thus chooses a different split $\alpha$ to achieve the target ESS of 30.
As a result, GMM-EnKF produces an intermediate distribution that is more tightly clustered on the observation in comparison to the other hybrids.
The posterior ensemble is also slightly less dispersed than the other hybrids, but is qualitatively similar.

To carefully compare the performance of the different methods, we solve the problem 1,000 times for each method over a range of ESS thresholds.
The results are compared on the basis of the root mean squared error (RMSE) where the mean is taken over the 1,000 experiments, and the continuous ranked probability score (CRPS; \cite{GR07,H00}).
The median of these 1,000 CRPS values is used as a summary statistic.
We also run a standard SIR particle filter with $10^4$ particles, as a reference approximation of the true Bayesian posterior.

\begin{figure}
    \centering
    \includegraphics[width=\textwidth]{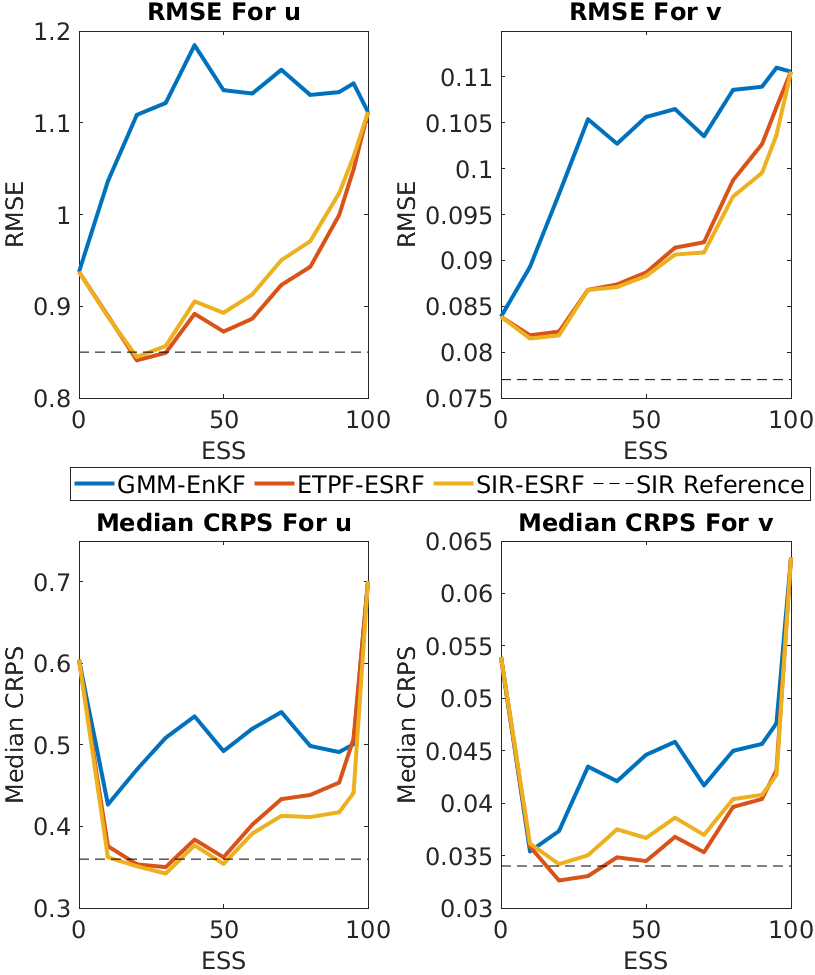}
    \caption{Filter performance over 1,000 trials as a function of ESS threshold. Top row: RMSE. Bottom row: Median CRPS. Left column: Results for the $U$ variable. Right column: Results for the $V$ variable. The ETPF results are shown in the plots at ESS $=0$ even though the ESS for ETPF is typically around 4.4. The ESRF results are shown in the plots at ESS $=100$. The dashed line in each panel shows the performance of an SIR particle filter with $10^4$ particles.}
    \label{fig:Henon2}
\end{figure}

Figure \ref{fig:Henon2} shows the performance of the methods as a function of the ESS threshold.
The top panels show RMSE and the bottom panels show the median of CRPS for the $U$ (left) and $V$ (right) variables.
The mean ESS of the ETPF over 1,000 trials is 4.4, so the smallest ESS threshold was set to 10.
The ETPF performance is shown on the plots at ESS$=0$, purely for convenience.
The pure ESRF performance is shown on the plots at ESS$=100$.
In each panel the performance of the pure particle filter with an ensemble size of $10^4$ is shown for reference.

All three hybrids perform similarly, though the hybrid of \cite{FK13} performs slightly worse than the other two in terms of RMSE.
This may be because the first step of the GMM-EnKF hybrid uses a GMM whose component Gaussians all use a covariance matrix obtained from the full prior ensemble; performance might be improved by using a clustering approach in the GMM following \cite{BSN03}.
The hybrids of \cite{CRR16} and section \ref{sec:Hybrids} are both able to perform \emph{better} than the pure particle filter when the ESS threshold is low, and are able to nearly match the performance of the true Bayesian filter as approximated by the pure particle filter with $10^4$ particles.
This is because the pure particle filter with 100 particles is still limited by low ESS (as underscored by the typical ESS value of 4.4).

The differences in RMSE between the methods are fairly small -- on the order of 25\% at most.
Differences in CRPS, which measures the quality of the uncertainty quantification (UQ) associated with the ensemble, are much larger.
The hybrid filters all achieve nearly optimal UQ, achieving more than 50\% improvement in CRPS over both ETPF and ESRF.

The pure particle filter is quite general in the sense that it generates a consistent estimator of the true Bayesian posterior for a wide range of problems.
The cost of this generality is the requirement of a very large ensemble size.
The hybrids trade this generality for improved performance using smaller ensemble sizes on a specific subset of problems, namely those with medium nonlinearity.

The code and data associated with this section can be found in \cite{code}.

\section{Numerical experiment: Lorenz-`96}
\subsection{A two-scale Lorenz-`96 Model}\label{sec:L96}
The experiments in this section make use of a model inspired by the Lorenz-`96 model \cite{L96,L06} and developed in \cite{GL15}.
The standard two-scale (or `two-layer') Lorenz-`96 model includes two sets of variables, $X_k$ and $Y_{j,k}$.
There are fewer $X_k$ variables, and they evolve more slowly than the $Y_{j,k}$ variables, so the $X_k$ variables are typically viewed as `large-scale' while the $Y_{j,k}$ variables are viewed as `small-scale.'
The difficulty with this model is that it lacks a clear connection to a spatial field of a physical quantity like temperature or velocity, observations of which contain both large and small scales.
A model inspired by the Lorenz-'96 models that possesses a single set of variables $x_i$ with distinct large-scale and small-scale dynamics was developed in \cite{GL15} and has been used recently as a test model for data assimilation in \cite{Grooms20}.
The model is governed by a system of ordinary differential equations of the form
\begin{equation}
\dot{\bm{x}} = h\bm{N}_S(\bm{x}) + J\mathbf{T}^T\bm{N}_L(\mathbf{T}\bm{x}) -\bm{x} + F\bm{1}
\end{equation}
where $h,F\in\mathbb{R}$, $J\in\mathbb{N}$, $\bm{1}$ is a vector of ones, and
\begin{align}
\left(\bm{N}_S(\bm{x})\right)_i &= -x_{i+1}(x_{i+2}-x_{i-1})\\
\left(\bm{N}_L(\bm{X})\right)_k &= -X_{k-1}(X_{k-2}-X_{k+1}).
\end{align}
The number of state variables in $\bm{x}$ is $41J$; here $J=128$ for a total system dimension of 5248.
As in the Lorenz-`96 model, the indices extend periodically.
The matrix $\mathbf{T}$ projects onto the 41 largest-scale discrete Fourier modes and then evaluates that projection at 41 equally-spaced points on the grid of state variables.
The matrix $J\mathbf{T}^T$ interpolates a vector of length 41 back to the full dimension of $\bm{x}$.

The large-scale part of the model dynamics is obtained by applying $\mathbf{T}$ to $\bm{x}$.
The result is identical to large-scale dynamics of the standard Lorenz-`96 model, except that the large scales are coupled to small scales via the term $h\mathbf{T}\bm{N}_S(\bm{x})$.
While the Lorenz-`96 model is often configured with 40 large-scale variables (e.g. \cite{LE98}), \cite{GL15} used 41 variables so that the 20$^\text{th}$ Fourier mode is not split between large and small scales.
At small scales, the dynamics are the same as those of original Lorenz-`96 model but with the direction of indexing reversed.

\begin{figure}
    \centering
    \includegraphics[width=\textwidth]{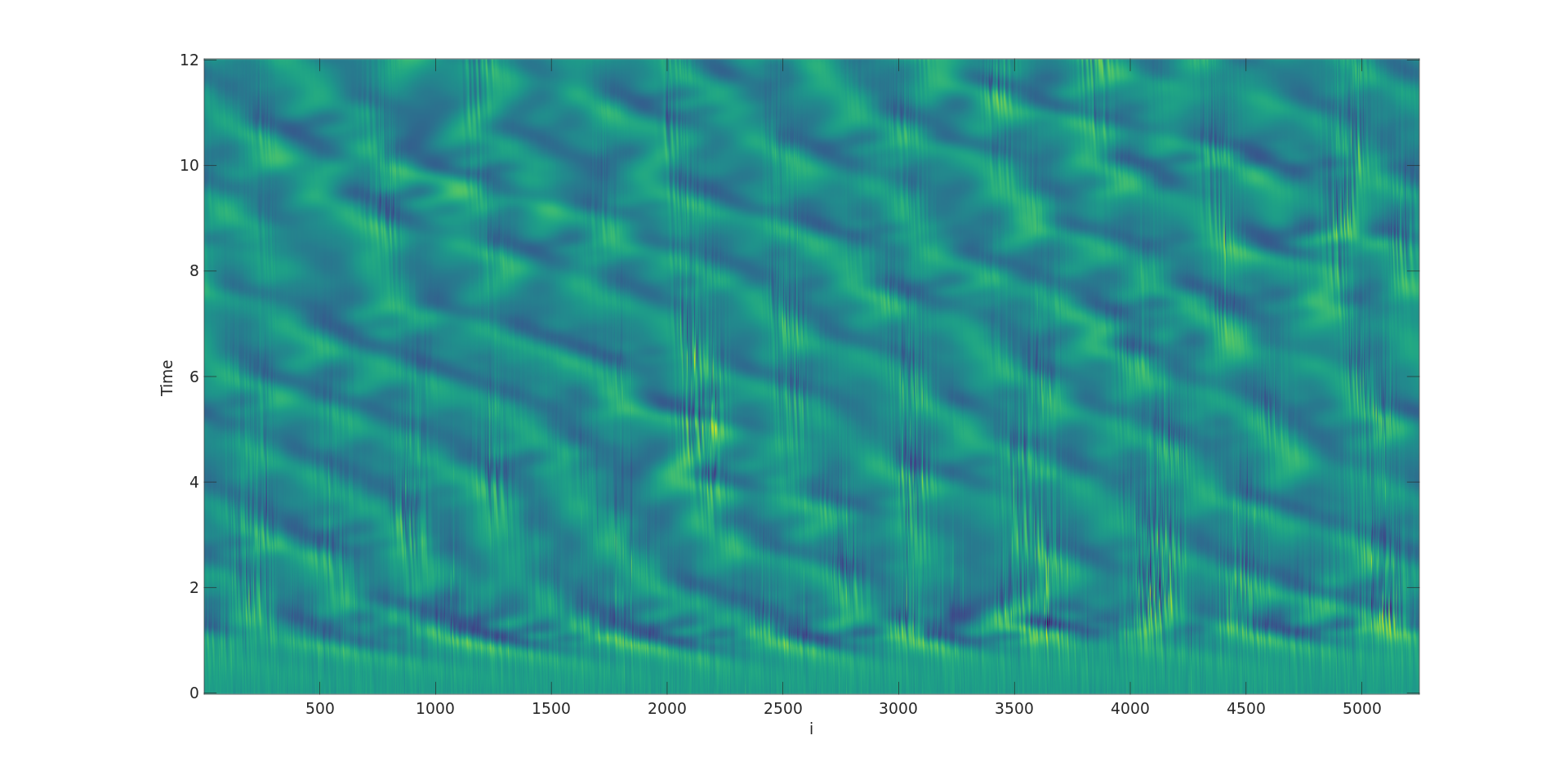}
    \caption{A simulation of the two-scale Lorenz-`96 model initialized at $t=0$ with a sample from a standard normal distribution.}
    \label{fig:L96}
\end{figure}

The experiments presented here use $h=0.38$ and $F=8$.
With these parameters the large-scale dynamics are very similar to the standard Lorenz-`96 model, with fairly weak coupling to the small scales.
The exception is when the large-scale Lorenz-`96 component reaches large values (e.g.~amplitudes $\ge 10$).
This occurrence excites a fast small-scale instability, causing the small scales also to reach large amplitudes that feed back locally onto the large-scale dynamics.
Fig \ref{fig:L96} shows the result of a simulation of this model initialized at $t=0$ with a sample from a standard normal distribution.
After a short transient the dynamics settle onto an attractor, with large-scale Lorenz-'96 modes propagating eastward and small-scale instabilities transiently excited by the large-scale waves.

\subsection{Data assimilation system configuration}\label{sec:DASetup}
\begin{figure}
    \centering
    \includegraphics[width=\textwidth]{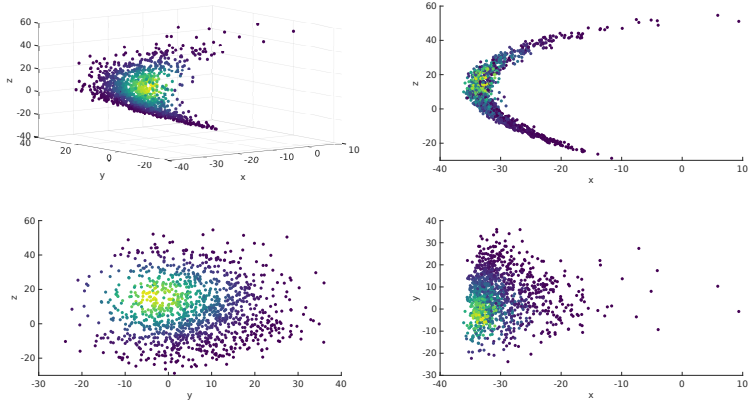}
    \caption{After one ensemble forecast (ensemble size is 1200) the deviations from the forecast mean are projected into the three leading eigenvectors of the empirical covariance matrix, with projection coefficients denoted $x$, $y$, and $z$. The four panels show four different perspectives on the projected ensemble. The color of each dot corresponds to the particle filter weight assigned using a split $\alpha$ chosen to yield an effective sample size of ESS = 600.}
    \label{fig:PointCloud}
\end{figure}

Reference solutions are generated by drawing initial conditions from an uncorrelated standard normal distribution and propagating the initial conditions by $9.0$ time units by numerical intergration of the dynamical model, at which point the state arrives at a statistical steady state (cf. Fig \ref{fig:L96}).
Upon reaching that statistically steady state, a reference state is produced at 1500 time intervals separated by 1.2 time units. 
In the usual interpretation of the standard Lorenz-`96 model, this time interval corresponds to 6 days, which is quite long compared to other studies.
At shorter time intervals the model exhibits only mild nonlinearity, where the forecast distribution is still very nearly Gaussian even though the dynamics are nonlinear.
At 6 days the forecast distributions are certifiably non-Gaussian, as shown in Fig \ref{fig:PointCloud}.
This figure was produced by projecting a forecast ensemble of 1200 members onto the three leading singular vectors of the ensemble's empirical covariance matrix.
The forecast distribution is dramatically non-Gaussian within this subspace --- therefore the EnKF assumption of a Gaussian prior is invalid.

Our hybrid is intended for situations with medium non-Gaussianity, where the prior is not Gaussian but the posterior is nearly Gaussian.
To achieve an approximately Gaussian posterior in the face of a non-Gaussian prior requires a large number of sufficiently-accurate observations.
Observations are taken at every fourth grid point (i.e. 32 observations for each of the 41 large-scale modes), with observation error variance $\gamma^2=1/2$.
This density and accuracy of observations is sufficient to produce a nearly-Gaussian posterior without rendering the data assimilation procedure superfluous.
(If the observations are dense enough and accurate enough then the filter adds essentially no information to the observations; this situation is avoided here, as the filter accuracy remains better than the observational accuracy.)

Ensemble members are initialized by propagating a sample from the uncorrelated multivariate standard normal distribution by 9.0 time units to arrive at an ensemble of substantially disparate states near the dynamic's attractor.
Because this initial forecast ensemble is fairly uninformative of the true state, there is a transient in filter performance while the filter  approaches its asymptotic optimal performance.
The results of the first 100 assimilation cycles are ignored in computations of filter performance statistics, so that the results presented are reflective of the statistical steady state of the filter.
The data assimilation system was run for 1500 cycles, i.e. nearly 25 years, for each trial in the experiment.

\subsection{Parameter Optimization}\label{sec:GP}
The ESRF used here has two primary tunable parameters: the inflation factor $r$ and the localization radius $L$.
The SIR-ESRF hybrid filter has an additional tunable parameter, the ESS threshold that determines the splitting factor $\alpha$.
The version of the hybrid filter with blurred observations (denoted BSIR-ESRF) also has tunable parameters related to the blurring, but these should not be viewed as a primary means of optimizing performance; we expect the hybrid to outperform the pure ESRF using only reasonable blurring parameters chosen a priori.
The demarcation between large and small scales occurs at Fourier wavenumber 20 for the Lorenz-`96 model considered here, so the blurring is chosen to have a Fourier spectrum
\[\frac{1}{\left(1+\left(\frac{k}{20}\right)^2\right)^2}.\]

To help substantiate a comparison between our SIR-ESRF hybrid approach and the pure ESRF filter, we independently tuned the respective filter parameters.
This began by generating parameter configurations, described hereafter as ``arms," from a Sobol sequence of low-discrepancy quasirandom numbers in a bounding box that we chose as a search space \cite{Owens98}.
(The term ``arm" comes from the literature on multi-armed bandits and denotes a particular configuration to be tested.)
The range of inflation factors considered was from $r=0$ to $r=0.08$ for the pure ESRF, and from $r=0$ to $r=0.15$ for the hybrid.
The range of ESS thresholds for the hybrid was from 66 to 400 for $N=400$ and 200 to 1200 for the $N=1200$.
The range of localization radius $L$ was from 128 points (equal to the separation between large-scale Lorenz-`96 modes) and 320 points.
At larger localization radii the filter performance became highly erratic, with some experiments performing extremely well and others extremely poorly.
It seems likely that at large localization radii there are rare occurrences of spurious long-range correlations that significantly degrade the filter performance.

For each arm, we ran at least four separate experiments with different reference solutions and initial ensembles.
For each assimilation cycle we computed the resulting root mean square error (RMSE), spread, and continuous ranked probability score (CRPS; \cite{GR07,H00}) for both the forecast (prior) and analysis (posterior).
RMSE and spread are scalar quantities at each timestep, but CRPS was computed for each state variable at each timestep.
We then aggregated these quantities by computing a mean over all state variables, timesteps, and assimilation trials ---
excluding the first 100 timesteps to allow for filter burn-in.

We elected to optimize for mean analysis CRPS because it quantifies the accuracy of the entire distributional estimate, whereas RMSE only describes accuracy of the ensemble mean point estimate.
The ensemble spread would also provide an estimate of the distributional accuracy, but CRPS is preferable in its ability to quantify the accuracy of non-Gaussian distributional estimates.
The median was excluded as an aggregation function to optimize because we found it to be insufficiently sensitive to situations in which the filter produces large intermittent excursion from the true state.

After exploring broad patterns with a Sobol sequence, we switched to a Bayesian optimization method for choosing new arms to evaluate.
Using a Bayesian optimization method substantially accelerated convergence to optimal filter parameters relative to the quasirandom search.
In short, this involved fitting a Gaussian process surrogate model to the mean CRPS observations as a function on the parameter search space, and then choosing new arms that maximize a utility function under that surrogate model.
We chose a utility function that estimates improvement from previously observed arms expected under the surrogate model.
The arms are then evaluated in parallel, by running the filter on a subset of the reference simulations using those arms' filtering parameters.
Those results are then incorporated with previous results to fit a new Gaussian process surrogate model used in the next iteration of the Bayesian optimization loop.
The technical details of the optimization strategy we used are described in \nameref{sec:Appendix}.

\section{Results: Lorenz-`96}
\label{sec:Results}
\begin{table}[]
    \centering
    \begin{tabular}{rrllll}
              &       & \multicolumn{2}{c}{Analysis} & \multicolumn{2}{c}{Forecast} \\
\cmidrule(lr){3-4} \cmidrule(lr){5-6}
        $N$ & Configuration & CRPS       & RMSE            & CRPS       & RMSE \\
\midrule
        400 & & & & & \\
        \cmidrule(lr){1-2}
            & ESRF  & 0.115 & 0.296 & 0.548 & 1.13 \\
            & SIR-ESRF  & 0.125 & 0.328 & 0.554 & 1.13 \\
            & BSIR-ESRF  & 0.111 & 0.287 & 0.533 & 1.10 \\
        1200 & & & & & \\
        \cmidrule(lr){1-2}
            & ESRF  & 0.112 & 0.280 & 0.539 & 1.11 \\
            & SIR-ESRF  & 0.115 & 0.281 & 0.530 & 1.08 \\
            & BSIR-ESRF  & 0.106 & 0.266 & 0.500 & 1.03 \\
    \end{tabular}
    \caption{Results for optimal parameter configurations of each method: pure ESRF, hybrid SIR-ESRF, and hybrid with blurred observations BSIR-ESRF. Results are averaged over the last 1400 assimilation cycles and over 8 different sets of initial conditions.}
    \label{tab:Results}
\end{table}

The three methods have indistinguishable performance at ensemble sizes smaller than 400, and the performance of all three methods improves with increasing $N$ up to $N=400$.
This suggests that for $N<400$ sampling errors limit filter performance more than errors due to non-Gaussianity.
It is probable that this threshold could be reduced with more sophisticated inflation and localization strategies (e.g. \cite{Anderson16,GRAW19}).
Table \ref{tab:Results} presents the results for the optimal parameter configurations of each method at two ensemble sizes $N=400$ and $N=1200$.

\subsection{$N=400$}
At an ensemble size of 400 the three methods yield very similar results.
In all three cases the filter is clearly doing better than simply trusting the observations, because the RMSE is nearly half the standard deviation of observation error.
The SIR-ESRF hybrid is slightly worse than the other two on average, because three of the eight runs produced significantly worse results, with analysis CRPS above 1.4.
In contrast, the BSIR-ESRF hybrid produces results quite similar to the ESRF.
One notable difference is that the optimal inflation parameter $r$ is larger for the hybrid filters than for the pure ESRF, presumably to counteract the under-dispersion that results from the resampling step in the particle filter.
(The optimal $r$ for SIR-ESRF is $0.06$ vs $0.026$ for ESRF.)
The optimal effective sample size for the SIR-ESRF hybrid was 297, which is fairly large compared to the ensemble size of 400.

\begin{figure}
\centering
   \includegraphics[width=1\linewidth]{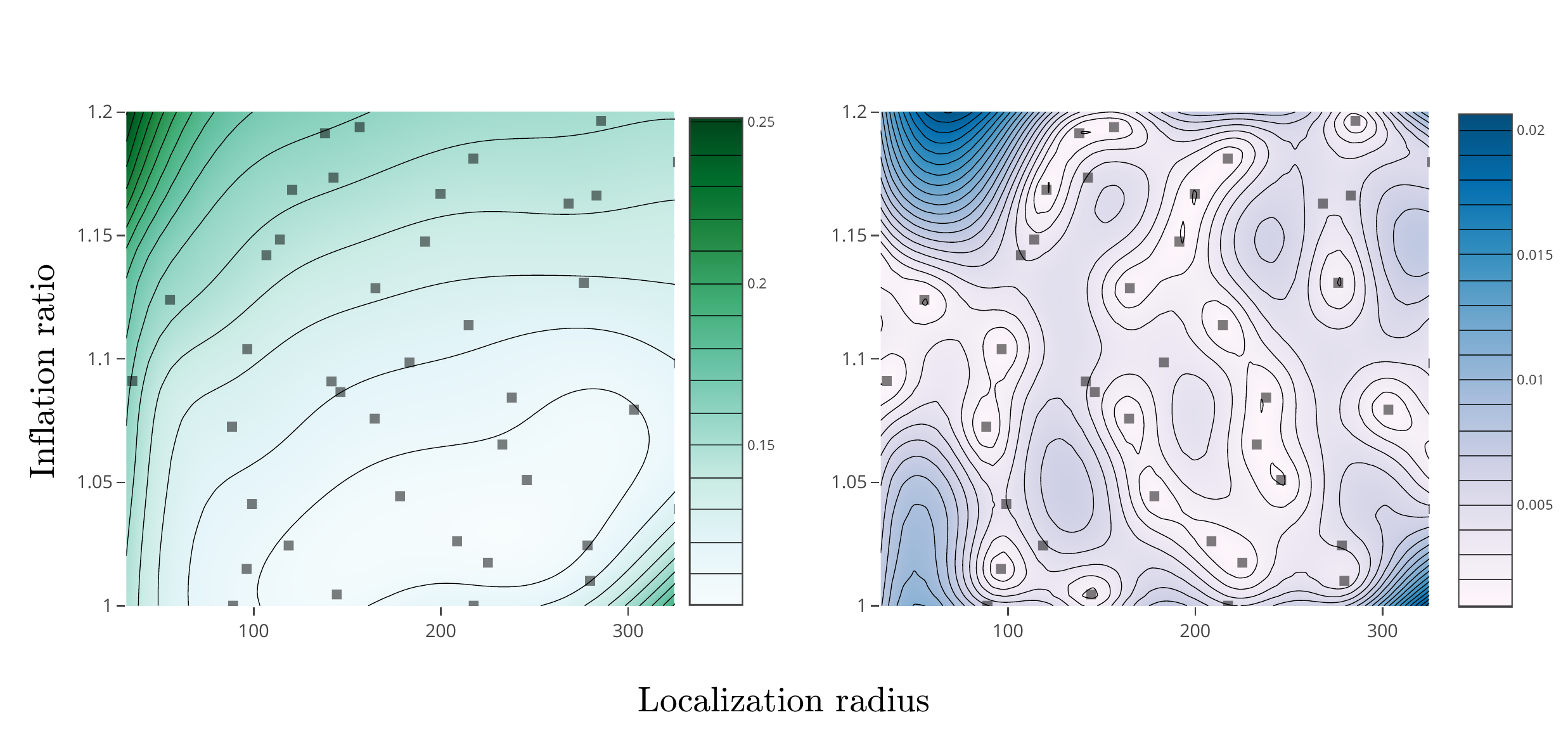}
   \caption{The Gaussian Process (GP) model of analysis CRPS for the pure ESRF model at $N=400$, showing analysis CRPS as a function of localization radius $L$ and inflation ratio $r$. The left panel shows the mean of the GP and the right shows the standard deviation. The small gray squares indicate parameter values where experiments were run.}
   \label{fig:400-analysis-crps} 
\end{figure}

Figure \ref{fig:400-analysis-crps} shows the GP surrogate model's prediction for analysis CRPS as a function of localization radius and inflation ratio ($1+r$) for the pure ESRF filter at $N=400$.
Gray squares indicate parameter configurations where experiments were run.
The left panel plots the mean of the GP, while the right panel plots the standard deviation.
The optimal parameters are in a fairly broad well, with near-optimal localization radii ranging from $100$ to $300$ and corresponding inflation factors from $r=0$ to $r=0.06$.
Interestingly, as the localization radius increases the corresponding optimal inflation factor does too.
At larger localization radii the filter makes more use of each observation leading to greater reduction in the posterior spread, which needs to be counterbalanced by increased inflation.
The GP surrogates for analysis RMSE and for forecast metrics are qualitatively similar, as is the behavior of the hybrid filters.
The optimal localization radii for the three filters are $L=209$ (ESRF), $L=279$ (SIR-ESRF), and $L=238$ (BSIR-ESRF).
These optimal values should not be over-interpreted, because the filters are not overly sensitive to the localization radius within the broad well that contains the optimal values.
Nevertheless, the fact that the hybrids are able to use a larger localization radius might suggest that the particle filter resampling step is eliminating outliers that would otherwise lead to spurious long-range correlations.

All of the methods lead to under-dispersed ensembles in the sense that the ensemble spread is less than the RMSE.
The forecast RMSE is 23\% larger than the forecast spread for both ESRF and BSIR-ESRF, while it is 30\% larger for SIR-ESRF.
Forecast spread is here measured before inflation, but in every case the inflation is not enough to match the inflated spread to the forecast RMSE.
The under-dispersion is worse for the analysis ensembles, with RMSE bigger than spread by 50\% for ESRF and BSIR-ESRF, and by 80\% for SIR-ESRF.
This mismatch between spread and RMSE can be reduced by tuning the parameters (particularly by increasing the inflation), but only at the cost of increasing both the RMSE and the CRPS.

\subsection{$N=1200$}
When the ensemble size is increased from 400 to 1200 the performance of the pure ESRF remains essentially flat, with only a minor improvement in analysis RMSE.
This shows that for $N\ge400$ the performance of the pure ESRF is limited primarily by the Gaussian approximation rather than by sampling errors.
The optimal inflation parameter for ESRF reduces from $r=0.026$ at $N=400$ to $r=0.015$ at $N=1200$, and the optimal localization radius increases from $L=209$ to $L=250$.
This is consistent with the intuition that as ensemble size increases less inflation and localization are needed to counteract sampling errors.
The ensemble remains about as under-dispersed as at $N=400$, with forecast RMSE 22\% larger than forecast spread and analysis RMSE 42\% larger than analysis spread.

The performance of the hybrid filters improves with increased ensemble size, with small improvements in CRPS and RMSE.
The SIR-ESRF hybrid is now nearly indistinguishable from the pure ESRF, and the BSIR-ESRF hybrid outperforms both by 5-10\% in CRPS and RMSE.
It is not clear whether further improvements could be obtained by increasing the ensemble size, or whether the hybrids are already close to the true Bayesian posterior.
No investigations have been performed at larger ensemble sizes due to the computational expense of optimizing parameters with very large ensembles.

Blurring of the observations enables the BSIR-ESRF hybrid to slightly out-perform the ESRF and the SIR-ESRF hybrid.
The split parameter $\alpha$ for a fixed ESS threshold tends to be larger in the hybrid with blurred observations: at $N=400$ the median $\alpha$ for SIR-ESRF is $10^{-3.14}$ while the median $\alpha$ for BSIR-ESRF is $10^{-2.91}$; at $N=1200$ the median $\alpha$ for SIR-ESRF is $10^{-2.96}$ while the median $\alpha$ for BSIR-ESRF is $10^{-2.48}$.
This suggests that for a fixed split $\alpha$ the blurring increases the ESS, but when the split $\alpha$ is instead chosen to produce a desired ESS the smoothing instead impacts which ensemble members are selected for resampling.
Heuristically this can be explained as follows: The hybrid essentially decides a priori how many distinct ensemble members will remain after resampling (by fixing the ESS), so the only impact of the blurring will be on which ensemble members are eliminated and which are replicated.
The effect of blurring is to trade improved assimilation performance at large scales for degraded performance at small scales; this trade is effective because predictability on longer time horizons comes from the large scales.
Indeed, the RMSE of the analysis mean projected onto the large scale modes is better for BSIR-ESRF than the other two methods.

When the ensemble size increases from $N=400$ to $N=1200$ the optimal inflation for the SIR-ESRF hybrid decreases from $r=0.06$ to $r=0.04$, and the optimal localization radius increases from $L=279$ to $L=316$.
The optimal ESS threshold is 757, although results are not overly sensitive for ESS thresholds in the range of 500 to 800.
The optimal parameters of the BSIR-ESRF method are similar: $r=0.02$, $L=319$, and ESS threshold 642.
Code and summary data associated with this section can be found in \cite{code}.



\section{Conclusions}\label{sec:Conclusion}
This paper has developed a hybrid particle ensemble Kalman filter targeting applications with medium nonlinearity, i.e. applications where the prior (forecast) distribution is very non-Gaussian but the posterior (analysis) distribution is close to Gaussian.
It was argued in \cite{MH19} that variational methods are more appropriate than the EnKF in this situation, because they approximate the posterior as Gaussian whereas EnKF methods approximate the prior as Gaussian.
The hybrid developed here is a pure ensemble approach for problems with medium nonlinearity, not requiring any variational optimization.
The particle filter acts first and results in an intermediate distribution that is closer to Gaussian than the prior; this intermediate distribution is then presented to the EnKF, and matches more closely the Gaussian approximation inherent to the EnKF.
The hybrid developed here is similar in spirit to previously-developed hybrids (e.g. \cite{FK13,CRR16}).
The main differences, besides the emphasis on medium nonlinearity, are the use of a serial square root filter and of a random resampling of the posterior ensemble to break the particle degeneracy introduced by the resampling step of the particle filter.

The hybrid SIR-ESRF developed here includes a resampling step that reduces the number of distinct ensemble members seen by the EnKF part of the hybrid (which is, in this case, a serial square root ESRF).
The EnKF's performance is limited by sampling errors even in purely Gaussian problems, so reducing the number of distinct ensemble members used within the EnKF increases the sampling errors and can hurt performance.
Our hybrid is configured such that the ESS in the particle filter step is specified a priori, and we find that the optimal ESS threshold for the hybrid needs to be at least as large as the ensemble size needed to obtain optimal performance in the pure EnKF.
(ESRF performance stopped improving for ensemble sizes greater than 400, and the optimal ESS in the hybrid was between 500 and 800.)
This leads one to expect that a larger ensemble size is required for the hybrid to outperform a pure EnKF, so that the particle filter component of the hybrid can effectively resample from the full ensemble size down to a size that is still large enough to obtain good EnKF performance.
In problems where non-Gaussianity presents in the form of a few outliers in an otherwise nearly-Gaussian distribution, the hybrid will presumably need only a slightly larger ensemble size, so that it can eliminate outliers during the resampling step.
But in problems where the forecast exhibits pathological non-Gaussianities such as multi-modality or strong curvature such as that seen in Fig \ref{fig:PointCloud}, a much larger ensemble may be needed in order for the hybrid to outperform the pure EnKF.
The SIR-ESRF hybrid did not achieve significant improvements over the pure ESRF in our experiments on the multiscale Lorenz-'96 model, but the BSIR-ESRF using smoothing of the innovations achieved limited improvements (5-10\% improvement in CRPS and RMSE).
This limited improvement compared to a pure ESRF may be a reflection of the fact that non-Gaussianity of the forecast is confined to a fairly low dimensional subspace associated with the leading singular vectors, i.e. the fastest directions of expansion along the system's attractor.

\section*{Supporting information}
\paragraph*{S1 Appendix}\label{sec:Appendix}
Let $c_i(p)$ denote the observed mean CRPS for trial $i$ with parameters $p$.
It is reasonable to expect that the mean CRPS is a continuous latent function $f$ of the filter parameters for fixed values of observed data, initial ensembles, random resamplings, and random rotations.
But since these fixed values all vary in practice, we can view each $f$ as a realization of a random field $F$.
In this view, the quantities $c_i(p)$ are noisy observations of the random field's true mean $\overline{F}$.
Our Bayesian optimizer seeks the minimizer of $\overline{F}$ using these noisy observations.

Let $\overline{c}_p$ be the mean CRPS observed over all assimilation trials that were run with parameters $p$.
Then let $\sigma_{\overline{c}_p}$ be the empirical standard error of that mean, computed as the sample standard deviation divided by the square root of the number of trials.
For convenience in setting hyperparameters of the Gaussian process model, we scale the search space to the unit cube and standardize the observations.
The raw search spaces are hyperrectangles, so they are scaled in each coordinate in the obvious manner to arrive at a unit cube.
To standardize the observations, we subtract the mean of the set $\left\{ \overline{c}_p \right\}$, for all parameter sets $p$ previously evaluated in the experiment, and divide the result by the sample standard deviation $\sigma_{\overline{c}_p}$ of the same set.
The raw standard errors $\sigma_{\overline{c}_p}$ are simultaneously divided by $\sigma_{\overline{c}_p}$ to preserve their validity in this standardized output space.
We do not introduce new notation for these transformed quantities; the remainder of this section will treat $c$ in the standardized output space and will treat values of $p$ in the scaled parameter space.

In these scaled spaces, we form a surrogate model supposing that $f_p$ depends on $p$ as a Gaussian process
\begin{align}
    \mathcal{GP} &\sim \mathcal{N}(0, k(p,p')).
\end{align}
We take $k(p,p')$ to be the Mat\'ern covariance kernel
\begin{align}
    k(p_i,p_j) &= \frac{\Theta_s 2^{1-\nu}}{\Gamma(\nu)} \left( \sqrt{2\nu} d(p_i,p_j) \right) K_{\nu} \left( \sqrt{2 \nu} \cdot d(p_i,p_j) \right),
\end{align}
where $K_\nu$ is the modified Bessel function of the second kind, and
\begin{align}
    d(\mathbf{p}_i, \mathbf{p}_j) &= (\mathbf{p}_i - \mathbf{p}_j)^\top \mathbf{\Theta}_d^{-1} (\mathbf{p}_i - \mathbf{p}_j)
\end{align}
Here $\mathbf{\Theta}_d$ is a diagonal matrix of length scale hyperparameters.
Each of the scalars on the diagonal of $\mathbf{\Theta}_d$ corresponds to a length scale of a feature in the space of scaled filter parameters, and each is endowed with a Gamma distribution prior $\Gamma(\lambda_L, r_L)$ with shape $\lambda_L = 6$ and rate $r_L = 3$.
The factor $\Theta_s$ is another hyperparameter that controls the covariance function's overall scale, on which we also impose a Gamma distribution prior $\Gamma(\lambda_S, r_S)$ with shape $\lambda_S = 2$ and rate $r_S = 0.15$.
We let the smoothness parameter $\nu=5/2$ so that realizations are almost surely twice-differentiable.
Marginalizing over the latent function $f$ yields the posterior distribution with log density
\begin{align}
    \ln P \left( \mathbf{f} | \{ \mathbf{p}_i \}, \Theta \right)
        = &-\frac{1}{2} \mathbf{s}^\top (\mathbf{K} + \mathbf{\Xi})^{-1} \mathbf{s} - \frac{1}{2} \ln \left| \mathbf{K} + \mathbf{\Xi} \right| - \frac{N_p}{2} \ln(2\pi) \label{eq:log-likelihood} \\
            &+ \sum_{j=1}^{N_p} \left[ (\lambda_L - 1) \ln \left( \Theta_{L,j} \right)
                - r_L \Theta_{L,j} + \lambda_L \ln r_L - \ln \Gamma (\lambda_L)
            \right] \\
            &+ (\lambda_S - 1) \ln \left( \Theta_S \right)
                - r_S \Theta_{S,j} + \lambda_S \ln r_S - \ln \Gamma (\lambda_S),
\end{align}
where $K_{ij} = k(p_i,p_j)$ is a covariance matrix.
The equation above obtains by adding the log-likelihood of our hyperparameter priors to Equation 2.30 of \cite{WR06}.
The GP surrogate is then fit to the rescaled data by maximizing the log-density Eq \ref{eq:log-likelihood} using many restarts of the L-BFGS-B method \cite{BLNZ95} to arrive at the maximum a posteriori (MAP) estimator.
Finally, a batch of candidate arms is generated that approximately optimizes the batched noisy expected improvement acquisition function \cite{LKOB19} on the MAP estimator.
Batch sizes varied between 1 and 32 depending on computational resources available at the time.

\end{document}